\documentclass{iopjournal}

\usepackage{graphicx}
\usepackage{amsmath}
\usepackage{lineno}
\usepackage{mathrsfs}
\usepackage{xcolor}
\usepackage{siunitx}
\usepackage{dcolumn}
\usepackage{bm}
\usepackage{braket}
\usepackage{stix}
\usepackage{mathrsfs}

\usepackage{anyfontsize}
\usepackage{float}

\usepackage{xr}
\begin{document}

%\articletype{Article type} %	 e.g. Paper, Letter, Topical Review...

\title{Parity-mixing interference in laser-assisted photoionization}

\author{N. Ouahioune$^{1,*}$\orcid{0000-0002-1092-2821}, D. Hoff$^1$, P.K. Maroju$^{1}$\orcid{0009-0007-4000-4400}, C.L. Arnold$^{1}$, D. Busto$^{1}$\orcid{0000-0003-4311-3315}, A. L'Huillier$^{1}$\orcid{0000-0002-1335-4022}, M. Gisselbrecht$^{1}$\orcid{0000-0003-0257-7607} and S. Carlström$^{2}$\orcid{0000-0000-0000-0000}}

\affil{$^1$Department of Physics, Lund University, Lund, Sweden.}

\affil{$^2$Max-Born-Institut, Max-Born-Straße 2A, 12489 Berlin, Germany.}

%\affil{$^*$nedjma.ouahioune@fysik.lu.se}

\email{nedjma.ouahioune@fysik.lu.se}

%\keywords{sample term, sample term, sample term}

\begin{abstract}
Photoionization of atoms by high-order harmonics in the presence of a laser may lead to quantum interference from which information about the photoionization dynamics or the light fields can be extracted. Traditionally, this interference arises from two-photon transitions involving the absorption of consecutive harmonics combined with the absorption and stimulated emission of a laser photon. In this process, parity is conserved. Here, we investigate interference between one- and two-photon transitions in helium using high-order harmonics generated by a few-cycle laser and three-dimensional electron detection. In this case, parity is not conserved. We identify four parity-mixing interference pathways, involving two different harmonic fields or a single harmonic, together with absorption or emission of a probe photon.
\end{abstract}

\section{Introduction}
Laser-assisted photoionization, which consists in ionizing matter with extreme ultraviolet (XUV) radiation in the presence of a dressing laser field \cite{SchinsPRL_1994,GloverPRL_1996}, has become a cornerstone technique in attosecond science. It has been used in particular to characterize the temporal properties of attosecond pulses, either in a train through the reconstruction of attosecond beating by interference of two-photon transitions (RABBIT) technique or for isolated pulses with the streaking method \cite{SansoneScience_2006,HentschelNature_2001,GoulielmakisScience_2008,PaulScience_2001,LopezMartensPRL_2005}. 

\begin{figure}[ht]
    \centering
    \includegraphics[width=0.55\linewidth]{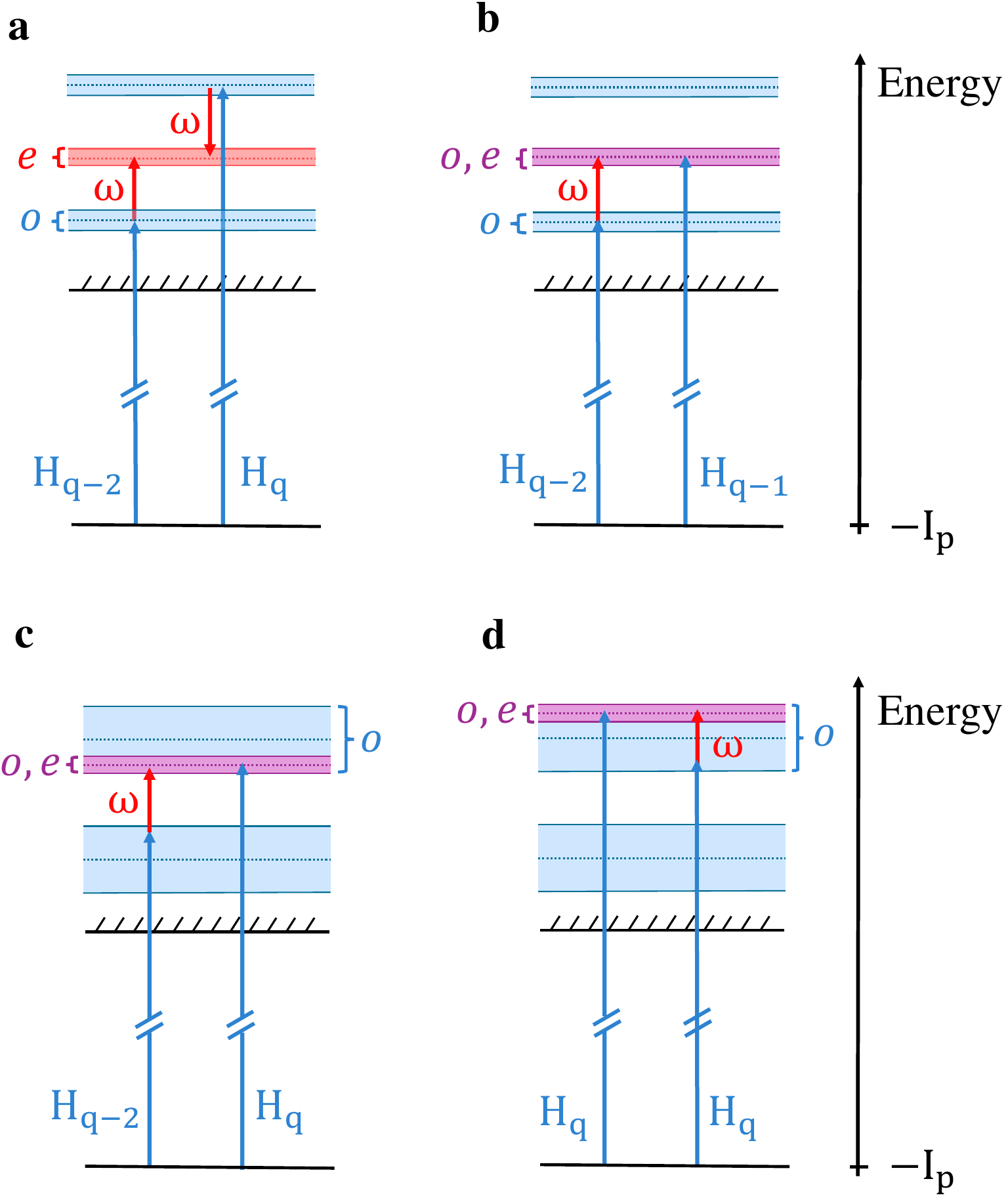}
    \caption{Energy diagram illustrating the (a) RABBIT technique and (b-d) techniques that lead to parity mixing interference in the case of IR absorption (similar schemes exist for IR emission). The blue (red) shaded area represents the mainband (sideband) spectral range. A purple color indicates spectral overlap. (b) RABBIT with odd and even harmonics, (c,d) RABBIT with broad light fields. In (c), the interference involves EWPs created by two different harmonics, while in (d), only one harmonic is involved. For simplicity, we only show the spectral range of the sideband that overlaps with the main band. The parity of the final state, odd ($o$) or even ($e$), is indicated in the figure.}
    \label{fig:figure_1}
\end{figure}

In RABBIT, a comb of odd high-order harmonics produces photoelectron wavepackets through ionization in the presence of a weak infrared (IR) laser field. The photoelectron spectrum consists of main bands due to absorption of the harmonics, and of sidebands due to the additional absorption or emission of a laser photon, as shown in Fig. \ref{fig:figure_1}a. The photoelectron yield oscillates as a function of the delay between the XUV and IR fields due to interference involving quantum paths conserving parity. The phase of the oscillations encodes temporal information about the attosecond pulses \cite{MairesseScience_2003} and the photoionization dynamics \cite{KlunderPRL_2011,IsingerScience_2017}. In streaking, a single attosecond pulse produces a broadband EWP in the presence of a strong IR field, whose energy oscillates with delay. Information about the XUV pulse or the photoionization process can be extracted from this time-energy mapping \cite{SchultzeNJP_2007, GoulielmakisScience_2008}.

Other techniques based on interference between quantum paths of different parities, as exemplified in Fig.~\ref{fig:figure_1}b-d, have also been proposed. This type of interference can be achieved in different ways. Odd and even harmonics can be generated by driving high-order harmonic generation (HHG) with the fundamental field and its second harmonic leading to parity mixing interferences, as shown in Fig.~\ref{fig:figure_1}b \cite{Laurent2012_PRL}. A similar scheme has been reported by using the second harmonic as the dressing field in the photoionization process, leading to spectral overlap between main- and sidebands \cite{LoriotJPhysB_2020}. Recently, parity-mixing interferences have been observed by using short laser pulses so that the side- and main bands overlap, as shown in  Fig.~\ref{fig:figure_1}c,d \cite{Fuch2021_PRR}. Finally, similar schemes have been explored at free-electron lasers \cite{PrinceNatPhot_2016, YouPRX_2020, MarojuNat_2020}. The observation of parity-mixing interference requires an electron detection that breaks the spatial symmetry. Therefore, most of the experiments mentioned above use momentum imaging techniques \cite{EppinkRSI_1997,UllrichRPP_2003}.

In this article, we perform laser-assisted photoionization of helium using harmonics generated by an ultrashort laser pulse, enabling spectral overlap between main- and sidebands, and three-dimensional electron-momentum detection. We identify four different parity-mixing interference pathways, involving the same harmonic (Fig.~\ref{fig:figure_1}d) and consecutive harmonics (Fig.~\ref{fig:figure_1}c). These overlapping pathways can be disentangled by Fourier analysis of the electron oscillations as a function of the XUV-IR delay. The phase and amplitude of the interferences encode information about the light fields and photoionization dynamics. Our results are supported by theoretical calculations based on the solution of the time-dependent Schrödinger equation. 
%Finally, we show that some of these spectral interferences are a direct consequence of the temporal interference between pairs of attosecond pulses, while interferences involving at least three attosecond pulses are never observed. 

\section{Experimental method}
\begin{figure}[ht]
    \centering
    \includegraphics[scale=0.75]{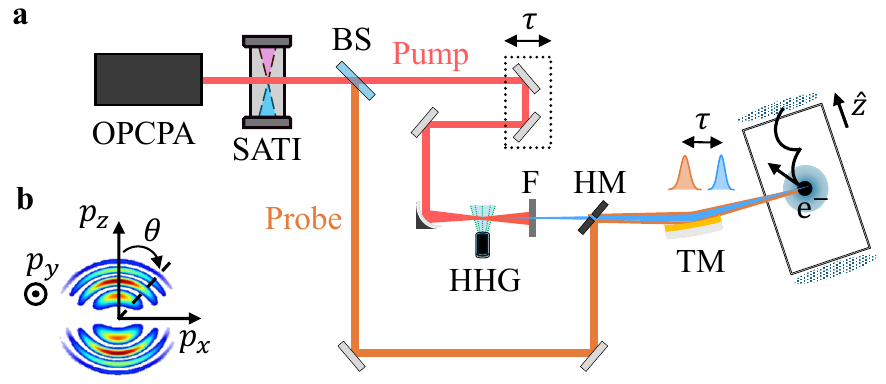}
    \caption{(a) Schematic of the experimental setup. OPCPA: Optical Parametric Chirped Pulse Amplification. SATI: Stereo-Above-Threshold-Ionization. BS: Beamsplitter. F: (Aluminum) Filter. HM: Holey Mirror. TM: Toroidal Mirror. (b) Three-dimensional momentum distribution of the photoelectron counts.}
    \label{fig:figure_2}
\end{figure}

We perform RABBIT measurements using the pump-probe interferometer shown in Fig.~\ref{fig:figure_2} \cite{MikaelssonJNanophotonics_2021}. An Optical Parametric Chirped Pulse Amplification (OPCPA) system operating at 200 kHz delivers few-cycle ($\sim$6 fs) pulses centered at 830 nm. The carrier-envelope-phase (CEP) of the pulses is stabilized using a stereo-above-threshold-ionization phase meter (SATI) \cite{WittmannNatPhys_2009}. The pulses are split into a pump (red) and a probe (orange) arm. The delay $\tau$ between the pump and probe is varied using a piezo stage in the pump arm. HHG in argon generates odd harmonics of the driving laser frequency. At the output of the gas jet, a 200-nm-thick aluminum filter (F) transmits only the XUV, which is then recombined with the IR probe using a holey mirror (HM). The co-propagating XUV pump and IR probe are finally focused by a toroidal mirror (TM) into a vacuum chamber containing helium gas. The three-dimensional momentum of the photoelectrons is measured with a reaction microscope \cite{GisselbrechtRevSciInst_2005}, as illustrated in Fig.~\ref{fig:figure_2}b. 

\begin{figure}[ht]
    \centering
    \includegraphics[width=.45\linewidth]{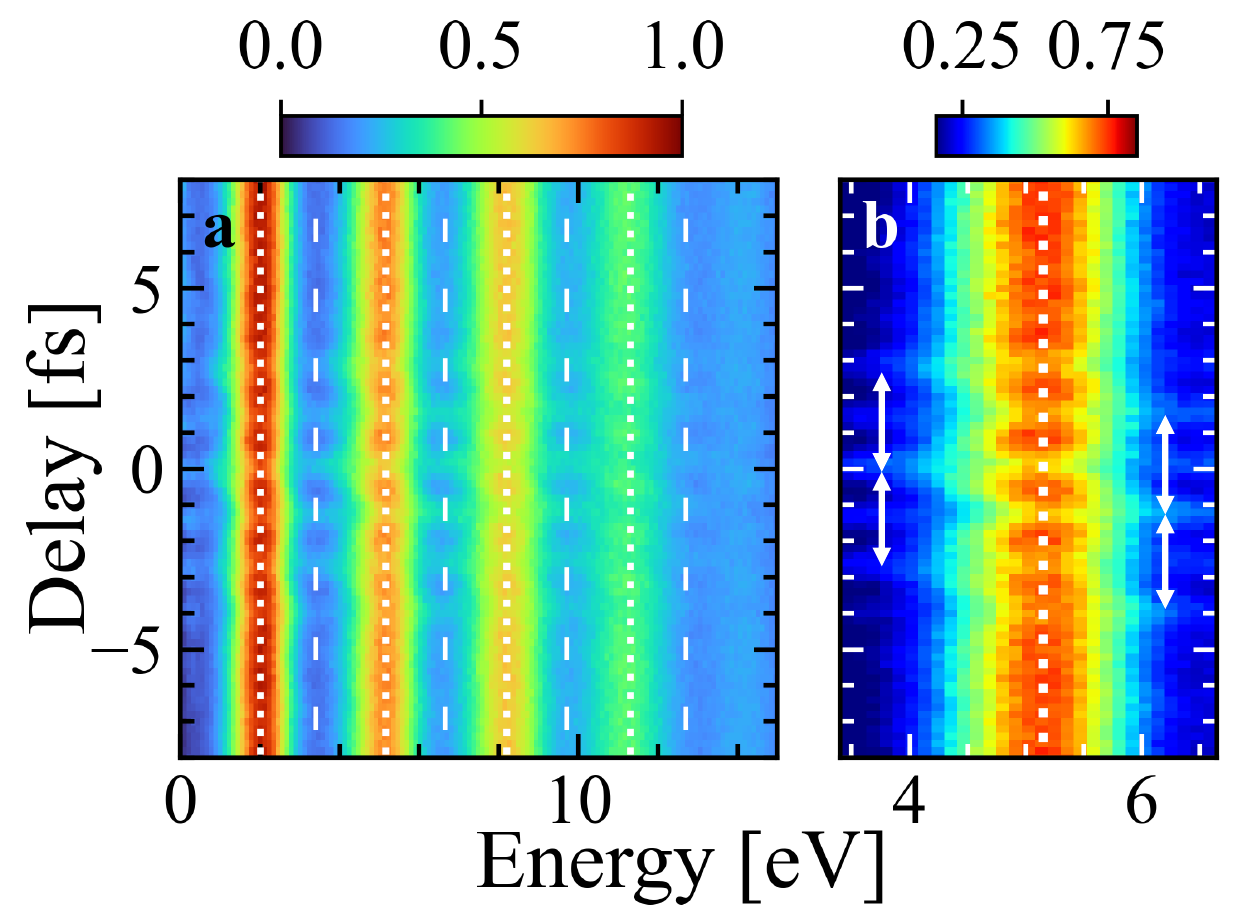}
    \caption{(a) RABBIT spectrogram showing the photoelectron counts as a function of kinetic energy and XUV-IR delay. Only electrons emitted in a 180$^\circ$ solid angle around the polarization axis of the light fields are included. White dotted (dashed) lines indicate the positions of the main (side) bands. (b) Spectrogram around the spectral region of the $\mathrm{ H_{19} }$ main band, which also shows oscillations at $\omega_0$ (white arrows).}
    \label{fig:figure_3}
\end{figure}

Fig.~\ref{fig:figure_3}a shows the signal (in color) of the photoelectrons emitted in the upper hemisphere of the detector, with momentum $p_z>0$, as a function of the XUV-IR delay. At long delays, the XUV and IR pulses do not temporally overlap. Broad photoelectron peaks are visible, corresponding to photoionization by absorption of the harmonics 17, 19, 21 and 23, see white dotted lines. At temporal overlap ($|\tau|\leq4$ fs), sidebands oscillating at $2\omega_0$ appear between the main bands, see white vertical dashed lines. Similar oscillations are visible in the main bands with a $\pi$-phase shift, due to partial depletion of single ionization processes. 

Fig.~\ref{fig:figure_3}b focuses on the spectral region between 3.5 and 6.5 eV. The main feature is due to absorption of harmonic 19. On each side, consecutive sideband peaks have different widths and amplitudes that oscillate out of phase. The white arrows point to the sideband peaks with the maximum amplitude. In this region, the maxima of the main band signal are slightly tilted spectrally. 

%An additional modulation of the signal at $\omega$ can be observed, indicating parity-mixing interferences. The white arrows outline this modulation, which is out-of-phase between the low and high energy parts of the main band.

\section{Fourier analysis}
\begin{figure*}[ht]
    \centering
    \includegraphics[width=.9\linewidth]{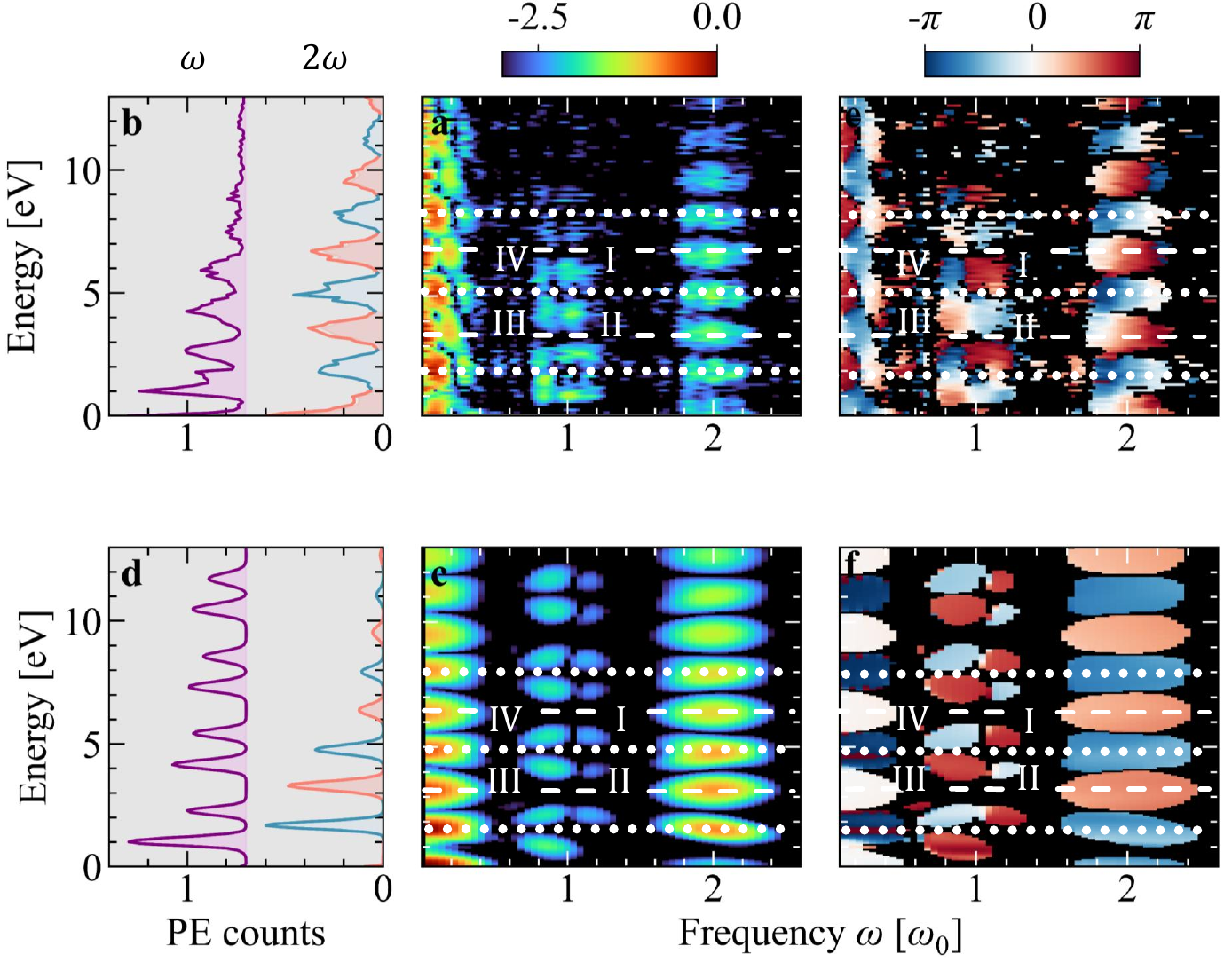}
    \caption{Amplitudes of the (a) experimental and (c) theoretical spectrogram Fourier transform. (b,d) Energy dependence of the oscillation frequencies in (a,c) arising from the interference between electrons in different ($\omega$) and in similar parity states ($2\omega$). Spectral phase of the (e) experimental (f) and theoretical data. Values below an arbitrary threshold have been removed (black) for the sake of clarity.}
    \label{fig:figure_45}
\end{figure*}

We perform a Fourier transform of the experimental spectrogram along the delay axis and show the resulting amplitude map in Fig.~\ref{fig:figure_45}a. Broad features around $2\omega_0$ and $\omega_0$ are visible. The former are strongest at the center of the main and sidebands (see white dotted and dashed lines), as expected from traditional RABBIT experiments. The latter, located between the white lines, exhibits two components, red- and blue-shifted relative to the central frequency $\omega_0$. 

Fig.~\ref{fig:figure_45}b shows the $2\omega$ signal for the main bands (blue) and for the sidebands (red) as well as the $\omega$ signal (purple). 
The purple curves overlap with the red and blue ones, indicating parity-mixing. Furthermore, the signal strength decreases towards the main- and sideband centers, where the spectral overlap is reduced. The $\omega$ signal at the lowest energy exhibits a double-peak structure, likely due to the effect of below-threshold Rydberg states \cite{MauritssonPRL_2010, LucchiniPRA_2015, ShivaramJPBAMOP_2016, RoantreePRA_2025}. Finally, the overall signal decreases with energy. Only the main bands corresponding to ionization by absorption of the harmonics 17 and 19 exhibit clear $\omega$ oscillations.

% \begin{figure}[ht]
%     \centering
%     \includegraphics[width=.7\linewidth]{Images_other/figure_4_norm.pdf}
%     \caption{Amplitudes of the (a) experimental and (c) theoretical spectrogram Fourier transform. (b,d) Energy dependence of the oscillation frequencies in (a,c) arising from the interference between electrons in different ($\omega$) and in similar parity states ($2\omega$).}
%     \label{fig:figure_4}
% \end{figure}

The experimental results are simulated by solving the three-dimensional time-dependent Schrödinger equation (TDSE)  within the configuration-interaction singles \emph{Ansatz} \cite{RohringerPRA_2006,CarlstromJPBAMOP_2018,CarlstromPRA_2022_I,CarlstromPRA_2022_II}. The spectral widths and intensities used for the harmonics are extracted from the experimental data, corrected by the helium ionization cross-section. For computational detail, see the supplementary information. Fig.~\ref{fig:figure_45}c shows the Fourier transform of the theoretical spectrogram. The main features observed in the experiment, in particular the splitting of the $\omega$ signal, are well reproduced. In the following, we concentrate on the interpretation of the $\omega$ signal features, indicated by the roman numbers in Fig.~\ref{fig:figure_45}a-c.

\section{Parity-mixing interference}
Each of the four $\omega$ substructures originates from interference that involve different spectral regions of the main- and sidebands, described in more details in the following. The substructures which are blueshifted relative to the central frequency $\omega_0$, labeled I and II, arise from interference involving neighboring main bands (``inter-harmonic'' interference). In I (II), a two-photon transition involving absorption of the $q+2$ ($q-2$) harmonic and emission (absorption) of a probe photon reaches the same final state as single ionization by absorption of the $q$ harmonic, as shown in Fig.~\ref{fig:figure_1}c for scheme II. The substructures which are redshifted relative to the central frequency $\omega_0$, labeled III and IV, arise from interference within the same main bands (``intra-harmonic'' interference). In III (IV), a two-photon transition involving absorption of the $q$ harmonic and absorption (emission) of a probe photon reaches the same final state as single ionization by absorption of the same harmonic ($q$), as shown in Fig.~\ref{fig:figure_1}d for scheme III.

\begin{figure}[ht]
    \centering
    \includegraphics[scale=0.3]{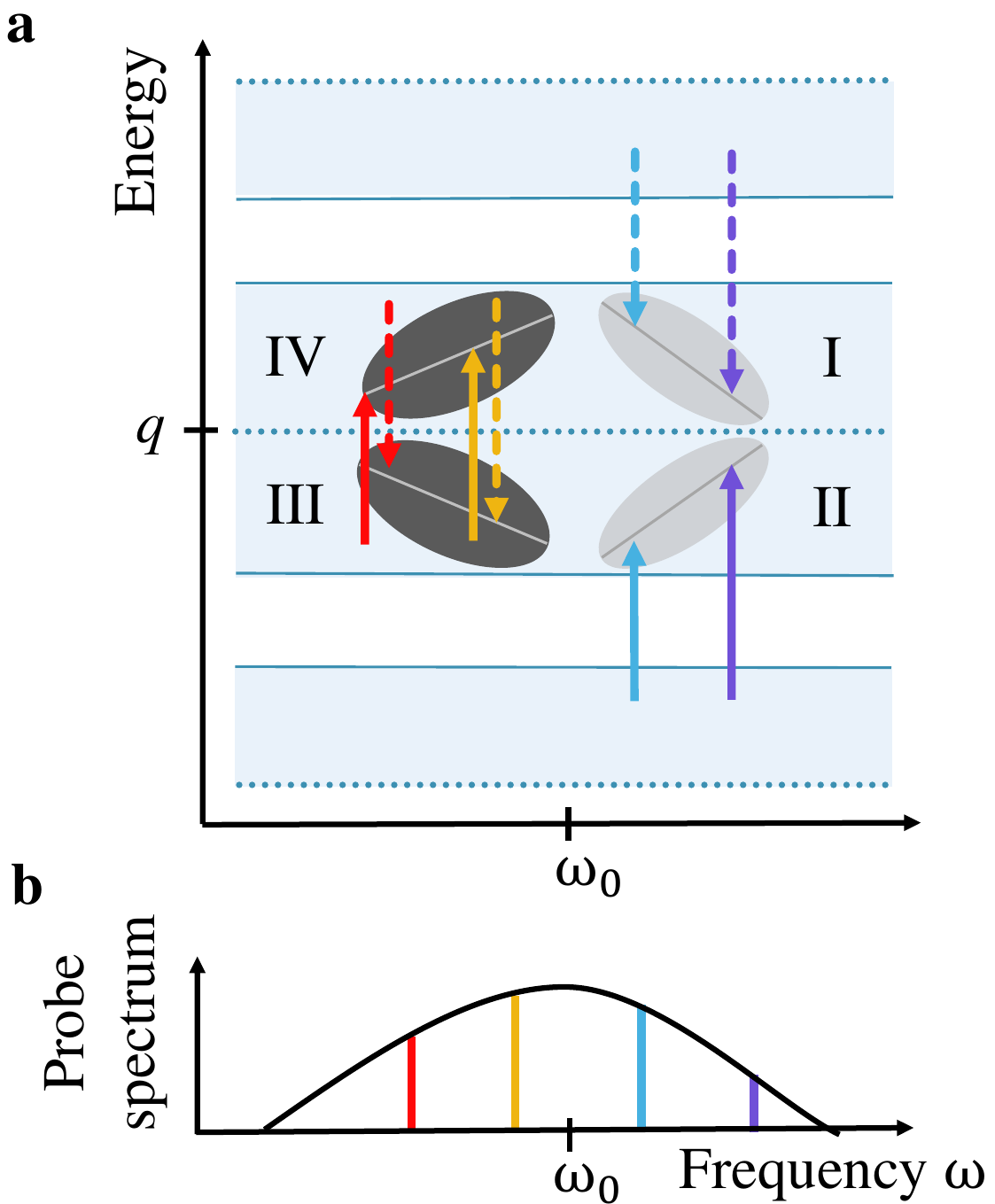}
    \caption{(a) Schematic illustrating interference pathways between final states of even and odd parity corresponding to ``inter-'' (I and II) and ``intra-'' harmonic (III and IV) interference. Interference involving different IR frequencies are shown with different colors, see probe spectrum in (b).}
    \label{fig:figure_6}
\end{figure}

Figure \ref{fig:figure_6}a illustrates these different interference pathways. Absorption (emission) of the probe photon are shown by solid (dashed) arrows for different frequencies of the probe spectrum, indicated by colors and illustrated in Fig.~\ref{fig:figure_6}b. Only the highest (lowest) probe frequencies lead to transitions I and II (III and IV), as shown by the colored arrows. A given final state is therefore primarily reached by a narrow range of probe frequencies, leading to a tilt of the interference pattern that is opposite in direction between emission and absorption, as well as between inter- and intra-harmonic interference. The tilt is small in the experimental data, due to the limited spectral resolution. However, it is visible in the theoretical TDSE data, as shown in Fig.~\ref{fig:figure_45}b. Finally, we present in Fig.~\ref{fig:figure_45}e,f the measured and simulated phase variations of the interference signal. Each substructure is surrounded by neighboring ones with a $\pi$ phase difference, which explains that the signal disappears at $\omega_0$ and in the middle of the main band. This structure arises because I and III involve the emission of a probe photon, whereas II and IV involve the absorption of a probe photon, as detailed below. 

In addition to the parity-mixing interference effects discussed above, Fig.~\ref{fig:figure_45}e-f shows phase variations at low ($< 0.4\omega_0$) and high frequencies ($\sim2\omega_0$) arising from the spectral variations of primarily the XUV field, which are not included in the simulations.

\section{Analysis of parity-mixing interference}
As illustrated in Fig.~\ref{fig:figure_1}c-d, the total signal at a given emission angle ($\theta,\phi$) and delay $\tau$ is the sum of interference pathways involving one- and two-photon processes, and can be written as
\begin{equation}
   \!\!\! S(\theta,\phi,\tau) \! = \!\! \bigg|\mathcal{A}^{(1)}_{ps}(\theta,\phi) + \!\! \!\!\sum_{\ell=0,2} \!\!\! \mathcal{A}^{(2,+)}_{\ell ps}(\theta,\phi,\tau) \! + \! \mathcal{A}^{(2,-)}_{\ell ps}(\theta,\phi,\tau)\bigg|^2\!\!\!,\!
    \label{eq:S_TDPT}
\end{equation}
where the index represents the angular momentum of the initial, intermediate and final states and where the exponent refers to the order of the process and to absorption (+) or emission (-) of the IR photon. The one- and two-photon transition amplitudes, $\mathcal{A}^{(1)}_{ps}$ and $\mathcal{A}^{(2,\pm)}_{\ell ps}$, are given by
\begin{align}
    &\mathcal{A}^{(1)}_{ps}(\theta,\phi)=a^{(1)}_{ps}Y_{10}(\theta,\phi),\\
    &\mathcal{A}^{(2,\pm)}_{\ell ps}(\theta,\phi,\tau)=a^{(\pm)}_{\ell ps}(\tau)Y_{\ell 0}(\theta,\phi),
\end{align}
with $Y_{\ell m}(\theta,\phi)$ being the spherical harmonics ($m=0$ in helium) and
\begin{align}
    &a^{(1)}_{ps}=\mathcal{E}_{q}(\Omega) M_{ps}^{(1)}, \label{eq:small_asp}\\
    &a^{(\pm)}_{\ell ps}(\tau) \!=\! \mathcal{E}_{\mathrm{IR}}(\omega) e^{\pm i\omega \tau}\! \left( \mathcal{E}_{q\mp2}(\Omega_\mp) M_{\ell ps}^{(2,\pm)} + \mathcal{E}_{q}(\Omega_\mp) M_{\ell ps}^{(2',\pm)} \right),
    \label{eq:small_a}
\end{align}
where $\Omega_\mp=\Omega\mp\omega$ and $\Omega$ are the intermediate and final state frequencies, $M_{ps}^{(1)}$ is the one-photon transition matrix element, $M_{\ell ps}^{(2,\pm)}$ is the two-photon transition matrix element involving different harmonics and $M_{\ell ps}^{(2',\pm)}$ is the two-photon transition matrix element involving only the q-th harmonic. $\mathcal{E}_{\mathrm{q}}$ ($\mathcal{E}_{\mathrm{IR}}$) is the complex amplitude of the q-th harmonic (IR) light field. 

From Eq.~(\ref{eq:S_TDPT}), the parity-mixing signal is 
\begin{equation}
    S_{\mathrm{PM}}(\theta,\phi,\tau) \! = 2\sum_{ \substack{\ell=0,2,\\\sigma=\pm} } \Re\{ \mathcal{A}^{(2,\sigma)}_{\ell ps}(\theta,\phi,\tau)\mathcal{A}^{(1)*}_{ps}(\theta,\phi)\},
   % \label{eq:S_PM_1}
\end{equation}
where we identify four contributions for each final orbital angular momentum corresponding to intra- and inter-harmonic processes [see Eq.~(\ref{eq:small_a})] discussed previously. 

Since we use linear polarized light, azimuthal symmetry arises, which lifts the dependence on $\phi$. Owing to the orthonormality of spherical harmonics, the parity-mixing feature vanishes upon integration over the full solid angle. As in the experiment, we integrate over the upper hemisphere
\begin{equation}
    \int_0^{2\pi} \int_0^{\pi/2} d\phi d\theta \sin(\theta) Y_{\ell0}(\theta,\phi)Y_{10}(\theta,\phi) = c_\ell,
\end{equation}
with $c_0 = \sqrt{3}/4 $ and $c_2 = \sqrt{15}/16 $. Due to the parity properties of the spherical harmonics, integration over the lower hemisphere changes the sign of the coefficients. The integrated parity-mixing signal reads
\begin{equation}
    S_{\mathrm{PM}}(\tau) \! = 2\sum_{ \substack{\ell=0,2,\\ \sigma=\pm} } c_\ell \Re\{ a^{(\sigma)}_{\ell ps}(\tau)a^{(1)*}_{ps}  \}.
    \label{eq:S_PM}
\end{equation}
The next step consists in substituting Eq.~(\ref{eq:small_asp}) and (\ref{eq:small_a}) 
\begin{equation}
    \begin{split}
        S_{\mathrm{PM}}(\tau) \! &=\!  2 \! \!  \sum_{ \substack{\ell=0,2} } \! \! c_\ell \Re\bigg \{ \! \!  \mathcal{E}_{\mathrm{IR}}^*(\omega) \mathcal{E}_{q+2}(\Omega_+) \mathcal{E}_{q}^*(\Omega) e^{- i\omega \tau} M_{\ell ps}^{(2,-) } M_{ps}^{(1)*} \! \!\! \!\! \! \\
        & + \mathcal{E}_{\mathrm{IR}}(\omega) \mathcal{E}_{q-2}(\Omega_-) \mathcal{E}_{q}^*(\Omega) e^{ i\omega \tau} M_{\ell ps}^{(2,+) } M_{ps}^{(1)*} \\
        &+  \mathcal{E}_{\mathrm{IR}}^*(\omega) \mathcal{E}_{q}(\Omega_+) \mathcal{E}_{q}^*(\Omega)  e^{- i\omega \tau} M_{\ell ps}^{(2',-)} M_{ps}^{(1)*} \\
        &+ \mathcal{E}_{\mathrm{IR}}(\omega) \mathcal{E}_{q}(\Omega_-) \mathcal{E}_{q}^*(\Omega)  e^{ i\omega \tau} M_{\ell ps}^{(2',+)} M_{ps}^{(1)*} \! \!  \bigg \} .
    \end{split}
    \label{eq:S_PM_int}
\end{equation}
In this expression, the four terms correspond to regions I, II, III and IV. In general, a $n$-photon transition amplitude is proportional to $(-i)^n$ \cite{Bertolino2021_PRR}. The different terms in Eq.~\ref{eq:S_PM_int} include the product of a two-photon matrix element ($\propto (-i)^2$) and the conjugate of a one-photon matrix element ($\propto i$). Neglecting small phase contributions to the fields and matrix elements, and using $\Re \{  i e^{\pm i\omega \tau} \}=\pm\sin{( \omega\tau)}$ we find that the different terms in Eq.~(\ref{eq:S_PM_int}) have alternating signs. This implies that the two terms corresponding to regions I and II or similarly III and IV are out-of-phase. The same argument applies to the two terms corresponding to regions II and III (or I and IV). This argument explains why the four regions in Fig.~\ref{fig:figure_45}e-f are $\pi$ out of phase and interfere destructively at the central probe frequency and the center of the main band.

\section{Conclusion}
In summary, we study laser-assisted photoionization in helium using a comb of broad harmonics and a few-cycle dressing field. We observe parity-mixing interference effects and identify four possible quantum pathways. A Fourier analysis allows us to spectrally distinguish these four quantum paths. In addition, a theoretical formalism based on perturbation theory shows that these quantum paths, which involve emission or absorption of a probe photon, are out of phase and interfere destructively when they spectrally overlap. Two of these quantum paths involve the same harmonic fields (intra-harmonic) while the other two involve different consecutive harmonics (inter-harmonic). As in a RABBIT measurement, the amplitude and phase of the parity-mixing interference encode information about the light fields and the one- and two-photon photoionization dynamics. With improved spectral resolution, this scheme opens the way to a complete reconstruction of the XUV and IR fields, as well as laser-assisted photoionization dynamics, including continuum-continuum transitions.

%In the case of broadband pulses, the variety of quantum paths allows for the unambiguous extraction of information about both the light fields and ionization dynamics, without requiring prior knowledge of either. Specifically, the phase of the light fields is directly encoded into the intra-harmonic interference signal while continuum-continuum phase differences and two-photon amplitude ratios between different scattering channels can be extracted from any of the parity-mixing substructures. This requires separating the four quantum paths, which implies high spectral resolution and, consequently, long pump-probe delays. 

\ack{S.~C. would like to thank Serguei Patchkovskii for helpful discussions.
}

\funding{M.~G., D.~B. and A.~L. acknowledge support from the Swedish Research Council (Grant Nos. 2020-05200, 2021-04691, 2023-04603, 2025-03729) and the Knut and Alice Wallenberg Foundation. A.~L. and D.~B. acknowledge support from the Knut and Alice Wallenberg Foundation through the Wallenberg Center for Quantum Technology (WACQT). A.~L. acknowledges support from the European Research Council (grants No. 884990, No. 851201).}

%\suppdata{Sample text inserted for demonstration.}

\suppdata{\textbf{Supplementary material.} Details about the TDSE simulations.}

%\section*{References}

\bibliography{bib.bib}

\end{document}